\documentclass[11pt, a4paper]{article}

\usepackage[utf8]{inputenc}
\usepackage[T1]{fontenc}
\usepackage{lmodern}
\usepackage{microtype}
\usepackage[pdfborder={0 0 0}]{hyperref}

\usepackage{color}
\usepackage{marginnote} 
\usepackage{ifthen} 
\usepackage{graphicx}

\usepackage[tbtags]{amsmath} 
\usepackage{amssymb}
\usepackage{amsthm}
\usepackage{dsfont} 
\usepackage{mathtools} 
\usepackage{tensor}

\usepackage[text={15.5cm, 23.0cm}, centering]{geometry}
\newcommand{\cif}{C^\infty}
\newcommand{\ee}[0]{\varepsilon}
\newcommand{\inv}[0]{{-1}}

\newcommand{\boldify}[1]{\ensuremath{\boldsymbol{#1}}}

\def\ba{\boldify{a}}

\def\be{\boldify{e}}

\def\bj{\boldify{j}}

\def\bm{\boldify{m}}

\def\bp{\boldify{p}}
\def\bq{\boldify{q}}

\def\bt{\boldify{t}}

\def\bx{\boldify{x}}
\def\by{\boldify{y}}

\newcommand{\CC}{\mathbb{C}}

\newcommand{\RR}{\mathbb{R}}

\newcommand{\algebra}[1]{\mathfrak{#1}}

\newcommand{\ah}{\algebra{h}}

\newcommand{\iso}{\algebra{iso}}
\newcommand{\so}{\algebra{so}}

\newcommand{\idadi}[1]{\bigl(\mathds{1}-\Ad(u_{#1}^{-1})\bigr)}

\newcommand{\defeq}{\coloneqq}

\newcommand{\isoeq}{\cong}

\newcommand{\weaklyequal}{\approx}

\newcommand{\diffd}{\mathrm{d}}

\newcommand{\tdiff}[2]{\frac{\diffd #1}{\diffd #2}}

\newcommand{\oo}{\otimes}
\newcommand{\surf}{\ensuremath{S_{g,n}}}

\newcommand{\grporbit}[1][i]{\mathcal{C}_#1}

\newcommand{\extphasespace}{\ensuremath{\mathcal{P}_{\mathrm{ext}}}}
\newcommand{\phasespace}{\ensuremath{\mathcal{P}}}
\newcommand{\nongaugephasespace}{\ensuremath{\pogr^{n-2+2g}}}

\newcommand{\csurface}{\ensuremath{\Sigma}}

\DeclareMathOperator{\Ad}{Ad}
\DeclareMathOperator{\diag}{diag}
\DeclareMathOperator{\Hom}{Hom}
\DeclareMathOperator{\ISO}{ISO}
\DeclareMathOperator{\Mat}{Mat}
\DeclareMathOperator{\PSL}{PSL}
\DeclareMathOperator{\SL}{SL}
\DeclareMathOperator{\SO}{SO}
\DeclareMathOperator{\Span}{span}
\DeclareMathOperator{\Tr}{Tr}

\newcommand{\loal}{\ensuremath{\so(2,1)}}
\newcommand{\poal}{\ensuremath{\iso(2,1)}}
\newcommand{\logr}{\ensuremath{\SO_+(2,1)}}
\newcommand{\pogr}{\ensuremath{\ISO(2,1)}}

\newcommand{\email}[1]{\href{mailto:#1}{\nolinkurl{#1}}}

\newtheorem{theorem}{Theorem}[section]

\newtheorem{lemma}[theorem]{Lemma}

\newtheorem{corollary}[theorem]{Corollary}

\newcommand{\ie}{i.e.\ }

\def\mytitle{Gauge fixing in (2+1)-gravity with vanishing cosmological constant}
\def\myauthors{C.~Meusburger and T.~Sch\"onfeld}
\hypersetup{pdftitle=\mytitle, pdfauthor=\myauthors}

\begin{document}

\begin{center}

{\huge\mytitle%
 \footnote{This work was funded by the research grant ME 3425/1-1 in the
   Emmy-Noether programme of the German Research Foundation (DFG).}}

\vspace{2em}

{\large
 C.~Meusburger\footnote{\email{catherine.meusburger@math.uni-erlangen.de}} \qquad\qquad
 T.~Sch\"onfeld\footnote{\email{torsten.schoenfeld@math.uni-erlangen.de}}}

\vspace{1em}

Department Mathematik, \\
Friedrich-Alexander-Universit\"at Erlangen-N\"urnberg \\
Cauerstra\ss e 11, 91058 Erlangen, Germany

\vspace{1em}

March 23, 2012

\vspace{2em}

\begin{abstract}
We apply Dirac's gauge fixing procedure to (2+1)-gravity with vanishing cosmological constant. For general gauge fixing conditions based on two point particles, this yields explicit expressions for the Dirac bracket. We explain how gauge fixing is related to the introduction of an observer into the theory and show that the   Dirac bracket is determined by a classical dynamical $r$-matrix. Its two dynamical variables correspond to the mass and spin of a cone   that describes the residual degrees of freedom of the spacetime. We show that different gauge fixing conditions and different choices of observers are related by dynamical Poincar\'e transformations. This allows us to locally classify all Dirac brackets resulting from the gauge fixing and to relate them to a set of particularly simple solutions associated with the centre-of-mass frame of the spacetime.
\end{abstract}

\end{center}

\section{Introduction}

(2+1)-dimensional gravity plays an important role in quantum gravity as a simple model for higher dimensions. It allows one to investigate important physics questions arising in the quantisation of the theory in a simplified framework, in which a full and rigorous quantisation of the theory is possible. This includes conceptual questions such as the role of time and observers in the quantum theory as well as mathematical questions surrounding quantisation, for an overview see \cite{Carlip:2003aa}.

In addition to its role in quantum gravity, (2+1)-gravity is of interest intrinsically due to its rich mathematical structure, which includes its close relation to Chern-Simons gauge theory and moduli spaces of flat connections \cite{Achucarro:1986aa, Witten:1988aa}, aspects of three-dimensional geometry \cite{Mess:2007aa, Benedetti:2005aa} as well as knot theory \cite{Witten:1989aa}, Reshetikhin-Turaev Invariants \cite{Reshetikhin:1991aa} and conformal field theory \cite{Witten:1989ab}.

While the quantisation of the Euclidean case is rather well understood, the quantisation of (2+1)-gravity with Lorentzian signature proves more difficult. This is due to the fact that (2+1)-gravity can be viewed as a constrained system and the implementation of the constraints in the quantum theory is achieved via representation-theoretical methods.  For (2+1)-gravity with Lorentzian signature, the relevant Lie groups and quantum groups whose representations arise in the construction of the quantum theory are non-compact, which causes difficulties that are not present in the compact cases.

This is a strong motivation to investigate approaches for the quantisation of this theory which implement the constraints directly into the classical theory via gauge fixing and then quantise the resulting classical description.  An independent motivation to study the effect of gauge fixing on the classical theory are indications that gauge fixing is related to the inclusion of an observer in the classical and quantum theory \cite{Meusburger:2009aa, Meusburger:2011aa}. The inclusion of observers into the quantum theory is an important conceptual question of quantum gravity, which can be studied in detail in this model.

A further motivation is the debate surrounding quantum group symmetries and non-commu\-tative structures in quantum gravity such as $\kappa$-Poincar\'e symmetries or Drinfel'd doubles \cite{Lukierski:1991aa, Lukierski:1992aa, Majid:1994aa, Kowalski-Glikman:2002aa, Freidel:2004aa, Meusburger:2009ab, Meusburger:2010aa}. While various models in quantum gravity exhibit quantum group symmetries, these quantum group symmetries are often associated with certain extended or enlarged Hilbert spaces, from which the gauge-invariant Hilbert space is obtained via the imposition of constraints. It is therefore unclear how much of this quantum group symmetry survives constraint implementation and if these symmetries are generic features of quantum gravity or merely technical tools in the construction of the quantum theory. It seems plausible that a model containing an observer would be useful in providing an answer to this question.

In this article, we summarise our results \cite{Meusburger:2011aa, Meusburger:2012aa} on gauge fixing in (2+1)-dimensional gravity with vanishing cosmological constant and discuss their physical interpretation and their implications in quantum gravity. We consider a  rather general set of gauge fixing conditions based on two point particles in the spacetime. These gauge fixing conditions have a direct physical interpretation as conditions that specify an observer in the spacetime. Via Dirac's gauge fixing procedure, we then derive an explicit description of the resulting Dirac bracket. We show that the underlying mathematical structures that define this bracket are classical dynamical $r$-matrices for the Lie algebra $\poal$. These classical dynamical $r$-matrices depend on two variables which have a direct interpretation as energy and angular momentum of the spacetime as measured by the associated observer.

We then discuss how different gauge fixing conditions are related by dynamical Poincar\'e transformations, which generalise the usual gauge transformations of classical dynamical $r$-matrices in the literature \cite{Etingof:1998aa}. We show that  these transformations  allow one to map each Dirac bracket  and the associated classical dynamical $r$-matrices to two simple standard solutions for almost all values of the dynamical variables. These standard solutions have a direct interpretation as the centre-of-mass frame of the spacetime.

Our article is structured as follows. In Section~2, we summarise the combinatorial description of the phase space of (2+1)-gravity which plays an important role in the quantisation of the theory and serves as the starting point for our gauge fixing procedure. We give a detailed discussion of its geometrical interpretation and show how the variables that parametrise the phase space describe the construction of spacetimes from regions in Minkowski space.

In Section~3 we show how this description of the phase space can be interpreted as a constrained system. We discuss the gauge fixing conditions imposed to obtain the gauge-invariant phase space of the theory and describe how gauge fixing the theory amounts to specifying an observer.  In Section~4, we determine the Dirac bracket associated with these gauge fixing conditions and show that it is given by classical dynamical $r$-matrices. We find that different choices of gauge fixing conditions are related by dynamical Poincar\'e transformations which depend on the total energy and angular momentum of the universe, as measured by this observer. We then discuss how these transformations allow one to obtain a centre-of-mass frame description of the spacetime, in which the centre of mass appears as an effective particle at rest at the origin or a ``tachyonic'' particle with a spacelike worldline. Section~5 contains our outlook and conclusions.

\section{The phase space of (2+1)-gravity}

\subsection{Notations and conventions}

We denote by $\be_0=(1,0,0)$, $\be_1=(0,1,0)$, $\be_2=(0,0,1)$  the standard basis of $\RR^3$ and use Einstein's summation convention.  Unless stated otherwise, all indices run from 0 to 2 and are raised and lowered with the three-dimensional Minkowski metric $\eta=\diag(1,-1,-1)$.  We write $\ee_{abc}$ for the totally antisymmetric tensor in three dimensions with  $\ee_{012}=1$. For three-vectors $\bx,\by\in\RR^3$, we write $\eta(\bx,\by)=\bx\cdot\by=\eta_{ab}x^ay^b$,  $\bx^2=\bx\cdot\bx$ and $\bx\wedge\by$ for the three-vector with components $(\bx\wedge\by)^a=\ee^{abc}x_by_c$.

The proper orthochronous Lorentz group in three dimensions is the group  $\logr \isoeq \PSL(2,\RR)$  with Lie algebra  $\loal \isoeq \algebra{sl}(2,\RR)$. In the following, we write $\Ad$ for its adjoint representation, which coincides with its representation by $\logr$ matrices.
The Poincar\'e group in three dimensions is the semidirect product of the proper orthochronous Lorentz group and $\RR^3$: $\pogr \equiv \logr \ltimes \RR^3$. We parametrise elements of $\pogr$ as
\begin{equation*}
  (u,\ba)=(u,0)\cdot (\mathds{1},-\bj)=(u, -\Ad(u)\bj)\quad
  \text{with}\; u\in \logr, \bj, \ba\in\RR^3,
\end{equation*}
and the group multiplication law then takes the form
$$
(u_1,\ba_1)\cdot (u_2,\ba_2)=(u_1\cdot u_2,\ba_1+\Ad(u_1)\ba_2).
$$
We fix a basis  $\{J_a, P_a\}_{a = 0, 1, 2}$ of the Lie algebra $\poal$, in which the Lie bracket takes the form
$$
[J_a, J_b] = \tensor{\ee}{_a_b^c} J_c,\quad [J_a, P_b] = \tensor{\ee}{_a_b^c} J_c,\quad [P_a, P_b] = 0.
$$
The basis elements $J_a$ are the generators of the Lorentz transformations, the basis elements $P_a$ the generators of the translations. An $\Ad$-invariant symmetric bilinear form on $\poal$, which defines the Chern-Simons formulation of (2+1)-gravity, is given by
\begin{equation}\label{eq:pairing}
\langle J_a,P_b\rangle=\eta_{ab},\quad \langle J_a,J_b\rangle=\langle P_a,P_b\rangle=0.
\end{equation}
We denote by $P_a^L, P_a^R$ the right- and left-invariant vector fields on $\pogr$ associated with the translations and by $J_a^L, J_a^R$ the ones associated with the Lorentz transformations.
More generally, for any basis
$\{T_a\}_{a=0,\dots,5}$ of $\poal$, we define
\begin{equation}\label{eq:vector-fields}
  L_a f(h)=\left.\tdiff{}{t}\right|_{t=0} f(e^{-tT_a} \cdot h),\quad
  R_a f(h)=\left.\tdiff{}{t}\right|_{t=0} f(h\cdot e^{tT_a})\qquad
  \forall f\in\cif(\pogr),
\end{equation}
where $e: \poal\to \pogr$, $\bx\mapsto e^{\bx}$ is the exponential map for $\pogr$.

\subsection{Phase space and Poisson structure}

In the following we investigate gauge fixing in (2+1)-gravity with vanishing cosmological constant in its formulation as a Chern-Simons gauge theory with gauge group $\pogr$ \cite{Witten:1988aa}. We consider spacetimes of topology $M\approx\RR\times\surf$, where $\surf$ is an oriented surface of genus $g$ with $n$ punctures representing massive point particles with spin.

It is shown in \cite{Witten:1988aa} that the physical or gauge-invariant phase space $\phasespace$ of $\pogr$-Chern-Simons theory on a manifold $M\approx\RR\times \surf$ is the moduli space of flat $\pogr$-connections on $\surf$.
This phase space can be parametrised in terms of group homomorphisms $h:\pi_1(\surf)\to\pogr$ from the fundamental group $\pi_1(\surf)$ into $\pogr$ that map the homotopy equivalence class of a loop $m_i$ around the $i$-th puncture to a fixed $\pogr$-conjugacy class $\grporbit$ determined by the mass $\mu_i\in[0,2\pi)$ and spin $s_i\in\RR$ of the associated particle \cite{deSousaGerbert:1990aa}:
\begin{equation*}
  h(m_i)\in\grporbit[i]=\{h\cdot \exp(-\mu_i J_0-s_iP_0)\cdot h^\inv \mid h\in \pogr\}.
\end{equation*}
Two group homomorphisms $h:\pi_1(\surf)\to \pogr$ describe the same physical state if and only if they are related by conjugation with $\pogr$. This implies that the gauge-invariant phase space of the theory is given by
\begin{align}\label{eq:phase-space}
  \phasespace
    &=\Hom_{\grporbit[1], \dots, \grporbit[n]}\big(\pi_1(\surf), \pogr\big)/\pogr\nonumber\\
    &=\{h: \pi_1(\surf)\rightarrow \pogr \mid h(m_i)\in\grporbit[i]\}/\pogr.
\end{align}

\begin{figure}
  \centering
  \includegraphics{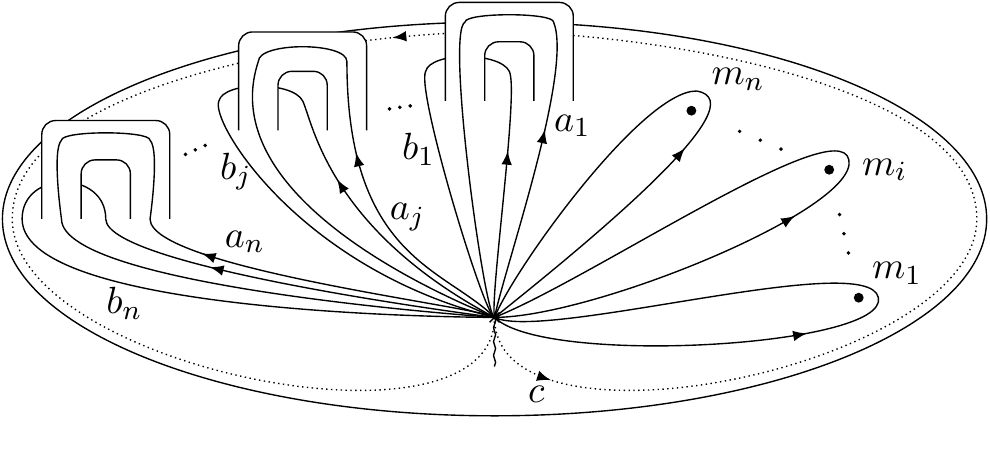}
  \caption{Generators of the fundamental group for an $n$-punctured genus $g$
    surface $\surf$.}
  \label{fig:fundamental-group}
\end{figure}

A set of generators of the fundamental group $\pi_1(\surf)$ is depicted in Figure~\ref{fig:fundamental-group}. It consists of a loop $m_i$ ($i=1,\dots,n$) around each puncture and the $a$- and $b$-cycles $a_j,b_j$ ($j=1,\dots,g$) for each handle. These generators are subject to a single defining relation which states that the curve $c$ in Figure~\ref{fig:fundamental-group} is contractible:
\begin{equation*}
  \pi_1(\surf)=\langle m_1,\dots,m_n, a_1,b_1,\dots,a_g,b_g \mid b_g a_g^\inv b_g^\inv a_g\cdots b_1 a_1^\inv b_1^\inv a_1 m_n\cdots m_1=1 \rangle.
\end{equation*}

By characterising group homomorphisms $h:\pi_1(\surf)\to\pogr$ in terms of the images $M_i=h(m_i)$, $A_j=h(a_j)$, $B_j=h(b_j)$ of the generators, one obtains a parametrisation of the phase space $\phasespace$ in terms of $\pogr$-matrices:
\begin{multline}\label{eq:phase-space-with-holonomies}
  \phasespace = \{(M_1,\dots,M_n,A_1,B_1,\dots,A_g,B_g)\in \pogr^{n+2g} \mid \\
                  M_i\in\grporbit, \  [B_g,A_g^\inv] \cdots [B_1,A_1^\inv] \cdot M_n\cdots M_1=1\}/\pogr.
\end{multline}
In the following, we refer to the images of elements $d\in\pi_1(\surf)$ under these group homomorphisms as holonomies and denote them by the associated capital letter $D=h(d)$ for $d\in\pi_1(\surf)$.

The moduli space $\phasespace$ of flat $\pogr$-connections is equipped with a symplectic structure that is induced by the canonical symplectic structure associated with the Chern-Simons action \cite{Atiyah:1983aa}.  A convenient description of this Poisson structure which serves as the starting point for its quantisation is derived in \cite{Fock:1998aa, Alekseev:1995ab}.  In this description, the Poisson structure on the moduli space is characterised in terms of an auxiliary, non-canonical Poisson structure on an extended phase space $\extphasespace=\pogr^{n+2g}$. This Poisson structure depends on the choice of a classical $r$-matrix $r=r^{\alpha\beta} T_\alpha\oo T_\beta\in\poal\oo\poal$ and is given in terms of a Poisson bivector:
\begin{align}\label{eq:fr-bivector}
 \{F,G\}&=B^r_{\text{FR}}(\diffd F \otimes \diffd G) \qquad\forall F,G\in\cif(\pogr^{n+2g}),\nonumber \\
 \begin{split}
    B^r_{\text{FR}}
    &=\tfrac 1 2 r_{(a)}^{\alpha\beta} \bigg(\sum_{i=1}^n L^{M_i}_\alpha+R^{M_i}_\alpha+\sum_{j=1}^g L^{A_j}_\alpha+R^{A_j}_\alpha+L^{B_j}_\alpha+R^{B_j}_\alpha\bigg)\oo\\[-.8em]
    &\qquad\qquad\qquad\bigg(\sum_{i=1}^n L^{M_i}_\beta+R^{M_i}_\beta+\sum_{j=1}^g L^{A_j}_\beta+R^{A_j}_\beta+L^{B_j}_\beta+R^{B_j}_\beta\bigg)\\[-.5em]
    &+ \tfrac 1 2 r_{(s)}^{\alpha\beta}\bigg( \sum_{i=1}^n R^{M_i}_\alpha\wedge R^{M_i}_\beta +\sum_{j=1}^g R^{A_j}_\alpha\wedge L^{A_j}_\beta+L^{B_j}_\alpha\wedge L^{B_j}_\beta\\[-.5em]
    &\qquad\qquad+\smashoperator{\sum_{1\leq i<j\leq n}} \left(L^{M_i}_\alpha+R^{M_i}_\alpha\right)\wedge\left(L^{M_j}_\beta+R^{M_j}_\beta\right)\\[-.5em]
    &\qquad\qquad+\sum_{i=1}^n\sum_{j=1}^g \left(L^{M_i}_\alpha+R^{M_i}_\alpha\right)\wedge\left(L^{A_j}_\beta+R^{A_j}_\beta+L^{B_j}_\beta+R^{B_j}_\beta\right)\\[-.5em]
    &\qquad\qquad+\smashoperator{\sum_{1\leq i<j\leq g}} \left(L^{A_i}_\alpha+R^{A_i}_\alpha+L^{B_i}_\alpha+R^{B_i}_\alpha\right)\wedge\left(L^{A_j}_\beta+R^{A_j}_\beta+L^{B_j}_\beta+R^{B_j}_\beta\right)\bigg),
  \end{split}
\end{align}
where $r^{\alpha\beta}_{(s)}=\tfrac 1 2 (r^{\alpha\beta}+r^{\beta\alpha})$, $r_{(a)}^{\alpha\beta}=\tfrac 1 2 (r^{\alpha\beta}-r^{\beta\alpha})$ denote the symmetric and antisymmetric components of $r$ and $L^X_\alpha, R_\alpha^X$ the right- and left-invariant vector fields \eqref{eq:vector-fields} associated with a basis $\{T_\alpha\}_{\alpha=0,...,5}$ of $\poal$ and with the different copies of $\pogr$.  It is shown in \cite{Fock:1998aa} that this bivector defines a Poisson structure on $\extphasespace$ if and only if $r$ is a classical $r$-matrix for $\poal$, \ie a solution of the classical Yang-Baxter equation
\begin{equation*}
  [r_{12}, r_{13}] + [r_{12}, r_{23}] + [r_{13}, r_{23}] = 0,
\end{equation*}
with $r_{12}=r^{\alpha\beta} T_\alpha\otimes T_\beta\otimes 1$, $r_{13}=r^{\alpha\beta} T_\alpha\otimes 1\otimes T_\beta$, $r_{23}=r^{\alpha\beta} 1\otimes T_\alpha\otimes T_\beta$. In this case, the resulting Poisson structure induces a symplectic structure on the moduli space $\phasespace$, which agrees with the canonical symplectic structure induced by the Chern-Simons action if and only if the symmetric part of $r$ is dual to the $\Ad$-invariant symmetric form \eqref{eq:pairing} in the Chern-Simons action:
\begin{equation*}
  r_S \equiv r_{(s)}^{\alpha\beta} T_\alpha\otimes T_\beta=\tfrac 1 2 (P_a\otimes J^a+J^a\otimes P_a).
\end{equation*}
In the application to (2+1)-gravity with vanishing cosmological constant, a natural  choice for the $r$-matrix which satisfies this condition is given by $r=P_a\oo J^a$. A detailed discussion of the Poisson structure on $\pogr^{n+2g}$ obtained from this choice of $r$-matrix is given in \cite{Meusburger:2003aa}, see also \cite{Meusburger:2010aa}.

\subsection{Geometrical interpretation}

Although the relation between $\pogr$-Chern Simons theory and (2+1)-gravity with vanishing cosmological constant is subtle \cite{Matschull:1999aa, Witten2007}, the link between the two theories is close enough to provide us with a direct geometrical interpretation of  the description of the gauge-invariant phase space $\phasespace$ in formula \eqref{eq:phase-space-with-holonomies}.

As the Ricci tensor of a three-dimensional manifold determines its curvature uniquely, vacuum solutions of the three-dimensional Einstein equations with vanishing cosmological constant are flat and locally isometric to Minkowski space. The theory has no local gravitational degrees of freedom and only a finite number of non-local degrees of freedom due to the presence of matter (point particles) and the  topology of the spacetime. These non-local degrees of freedom are encoded in  Poincar\'e transformations that describe the construction of spacetimes from  open regions in Minkowski space and are closely related to the holonomies  in the previous subsection.

\begin{figure}
  \centering
  \includegraphics{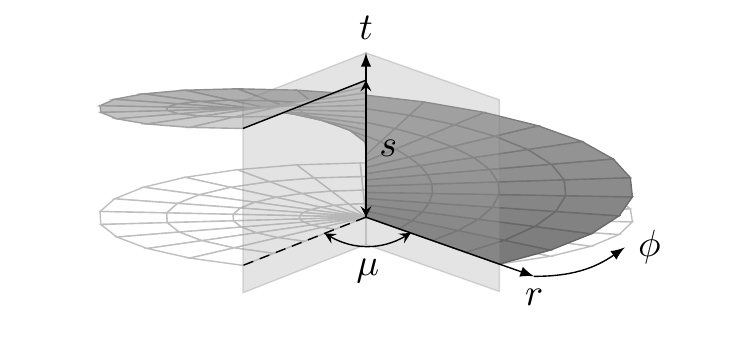}
  \caption{Conical spacetime associated with a point particle. The two horizontal black lines depict the identification of the boundary of the wedge according to $(t,r,0) \sim (t+ s, r, \mu)$.}
  \label{fig:wedge}
\end{figure}

The construction of spacetimes from  regions in Minkowski space is illustrated by the simplest example, namely the spacetime associated with a single point particle in Minkowski space.  The metric associated with a particle in three-dimensional Minkowski space that is at rest at the origin was derived in \cite{Deser:1984aa}, see \cite{Staruszkiewicz:1963aa} for earlier results on the Euclidean case. It is shown there that in cylindrical coordinates $(\tau,\rho,\phi)$ it takes the form
\begin{equation}\label{conem}
  \diffd s^2=(\diffd\tau+\tfrac s {2\pi}\,\diffd\phi)^2-\frac{1}{(1-\tfrac{\mu}{2\pi})^2}\diffd\rho^2-\rho^2 \diffd\phi^2,
\end{equation}
where $\mu\in[0,2\pi)$ is the mass and $s\in\RR$ the spin of the particle.  By introducing a new set of coordinates $(t, r, \varphi)$ that is related to the cylindrical coordinates $(\tau,\rho,\phi)$ via
\begin{equation*}
  t(\tau, \rho, \phi)=\tau+\tfrac s {2\pi}\, \phi,\qquad
  r(\tau, \rho, \phi)=\frac{\rho}{1-\tfrac\mu {2\pi}},\qquad
  \varphi(\tau, \rho, \phi)=(1-\tfrac\mu {2\pi})\phi,
\end{equation*}
one finds that the metric \eqref{conem} can be related to the Minkowski metric
\begin{equation*}
  \diffd s^2=\diffd t^2-\diffd r^2- r^2\diffd \varphi^2.
\end{equation*}
However, this metric is only locally, not globally, isometric to the Minkowski metric. This is apparent from the range of the variable $\varphi$, which is no longer $[0,2\pi)$ but $[0,2\pi -\mu)$. The metric thus describes a conical spacetime obtained by cutting out a wedge of Minkowski space and identifying its boundary according to $(t,r,0) \sim (t+ s, r, \mu)$ as shown in Figure~\ref{fig:wedge}. This identification is given by the action of the Poincar\'e transformation $M=(\exp(-\mu J_0), - se_0)$.

By applying a Poincar\'e transformation $(v,\bx)\in\pogr$, this description of the point particle metric is easily generalised to particles whose worldline is a general timelike geodesic in Minkowski space. Such a geodesic can be parametrised as $g(T)=T\,\hat \bp+\bx$ with $\bx\cdot \bp=0$, $\hat\bp^2=1$, where $T$ is the eigentime of the particle, $\hat\bp$ its unit momentum three-vector and $\bx$ the position of the particle at $T=0$. In this case, the identification of the boundaries of the wedge is given by a Poincar\'e transformation $M\in\pogr$ which takes the form
\begin{equation*}\label{mpartpar}
  M=(u,-\Ad(u)\bj)=(v,\bx)\cdot (\exp(-\mu J_0), -s\, e_0)\cdot (v,\bx)^\inv,\qquad
  u, v \in \logr,\;\bj, \bx \in \RR^3.
\end{equation*}
The relation between the
variables $u, \bj$ and $v, \bx$ is given by
\begin{equation}\label{ujexpress}
  u=v\cdot\exp(-\mu J_0)\cdot v^\inv=\exp(-\mu\, \hat p^c J_c), \qquad
  \bj=s\hat\bp+\idadi{}\bx.
\end{equation}
The quantity $\bp=\mu\hat\bp=\mu\Ad(v)e_0$ has the interpretation of a momentum three-vector of the particle, whose 0-component describes its energy and whose 1- and 2-component its momentum.  The quantity $\bj$ can be viewed as a generalised angular momentum three-vector. Its 0-component describes the angular momentum of the particle, its 2- and 3-component are related to Lorentz boosts. From the formula above it is apparent that the angular momentum three-vector is composed of two parts. The first one is parallel to its momentum three-vector and given by the particle's spin. The second one is orthogonal to the momentum three-vector and encodes the angular momentum of the particle due to its motion. In the limit $\mu\to0$, one has $\bj\to s\hat\bp+\hat\bp\wedge\bx$ and hence recovers the usual expression for the angular momentum.

\begin{figure}
  \centering
  \includegraphics[scale=0.7]{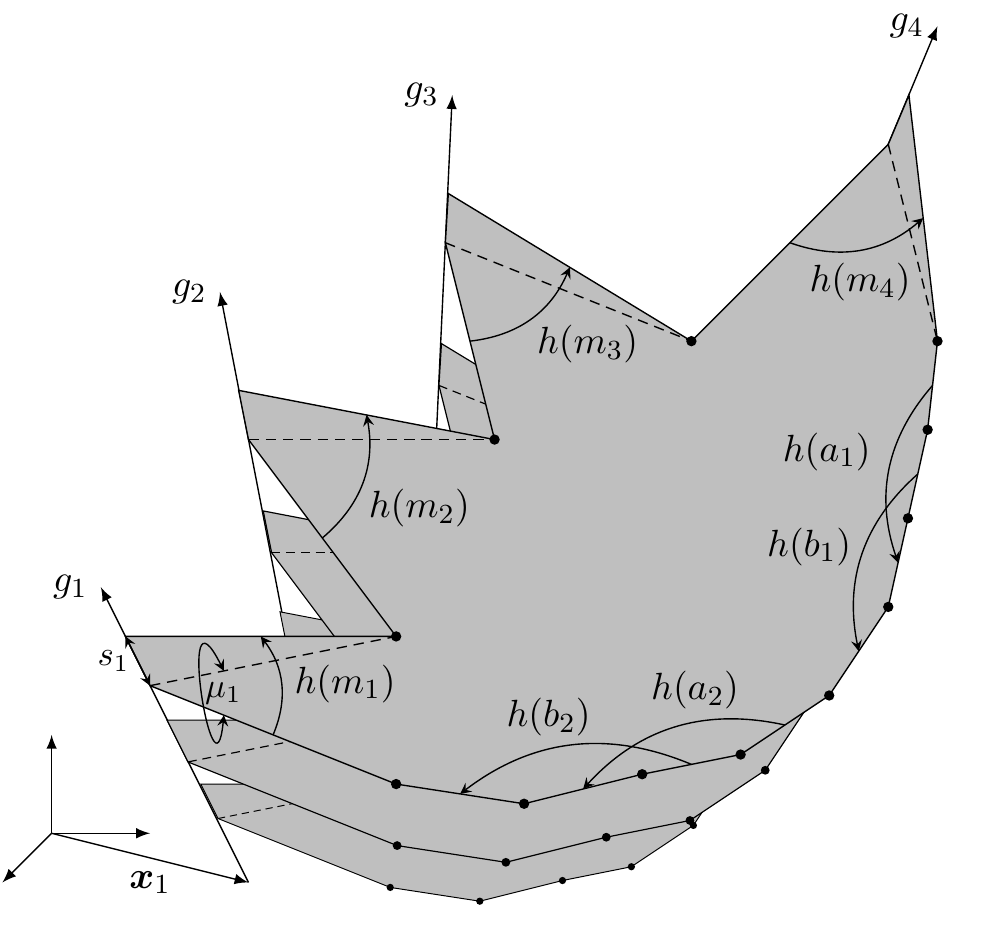}
  \caption{Domain in Minkowski space for a spacetime of genus $g=2$ with $n=4$ particles.}
  \label{fig:domain}
\end{figure}

\begin{figure}
  \centering
  \includegraphics[scale=0.7]{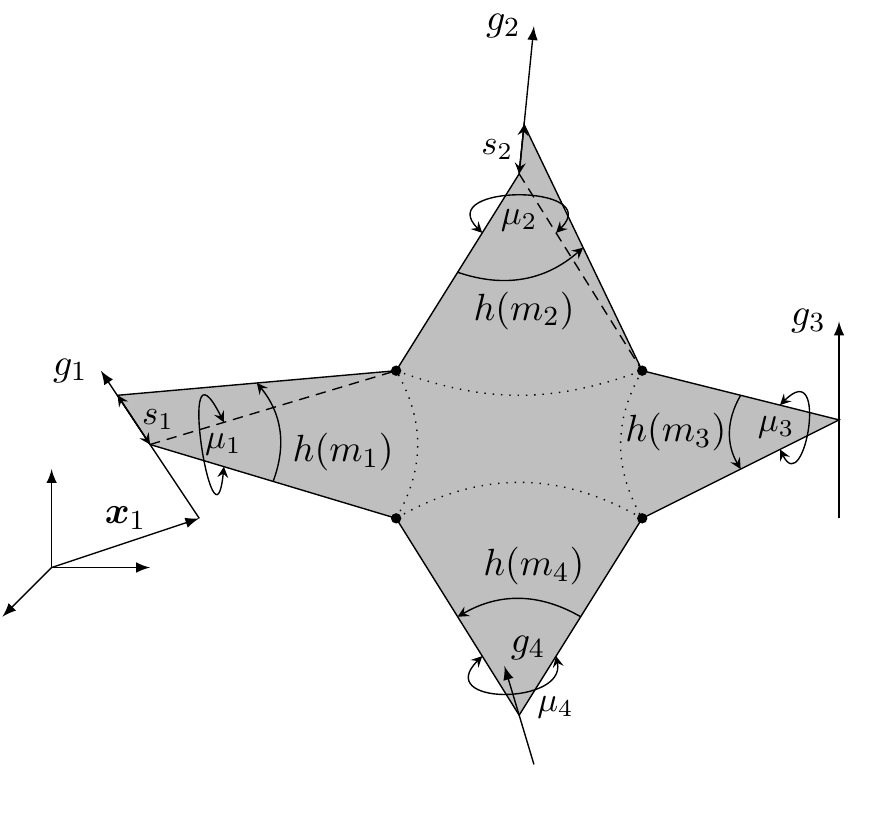}
  \raisebox{6.5em}{$\longrightarrow\;\;$}
  \raisebox{1em}{\includegraphics[scale=0.7]{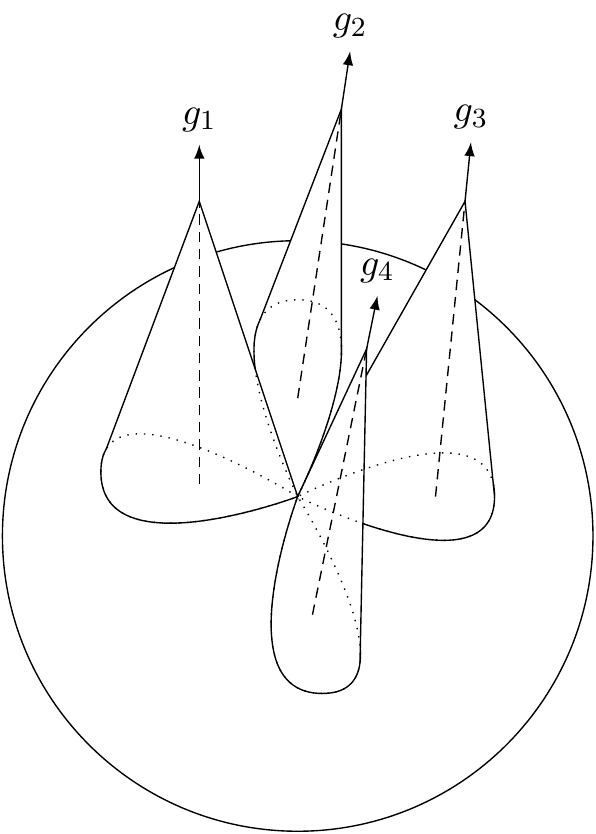}}
  \caption{Gluing of a spacetime of genus $g=0$ with $n=4$ point particles.}
  \label{fig:coneparticle}
\end{figure}

The general case of a spacetime $M\approx\RR\times\surf$ of genus $g$ with $n$ point particles is more involved, but the construction of spacetimes by identifying the boundary of an open region in Minkowski space is similar. The main difference is that the associated region in Minkowski space is no longer obtained by cutting out a wedge but takes a more complicated form depicted schematically in Figure~\ref{fig:domain}.  Its boundary is given by $2(n+2g)$ plane segments  in Minkowski space, which are identified pairwise by certain Poincar\'e transformations. These Poincar\'e transformations correspond to the images of the generators of the fundamental group $\pi_1(\surf)$ under the group homomorphisms $h:\pi_1(\surf)\to\pogr$ in formula \eqref{eq:phase-space-with-holonomies}. The associated spacetime is obtained by gluing these sides pairwise as shown in Figures \ref{fig:domain}, \ref{fig:coneparticle}.

The condition that the holonomies for the loops $m_i$ around the punctures lie in fixed conjugacy classes $\grporbit$ is needed to ensure that the identification of the sides corresponding to point particles gives rise to a cone with the correct opening angle and time shift defined by the mass and spin of the particle. The condition that arises from the defining relation of the fundamental group ensures that all black vertices in Figure~\ref{fig:domain} are mapped to a single point in the spacetime and guarantees that the gluing procedure gives rise to a flat Lorentzian manifold with conical singularities. The holonomies from the previous subsection thus have a direct interpretation as gluing data that describes the construction of spacetimes from regions in Minkowski space.

Given a region in three-dimensional Minkowski space whose sides are identified by the holo\-nomies $h(m_i), h(a_j), h(b_j)$, one obtains another such region by applying a Poincar\'e transformation $g\in\pogr$ to it. The sides of the image are then identified by the images of the holonomies under the diagonal action of $\pogr$ on $\grporbit[1]\times\ldots\times\grporbit[n]\times \pogr^{2g}$: $h(m_i)\mapsto g\cdot h(m_i) \cdot g^\inv$, $h(a_j)\mapsto g\cdot h(a_j) \cdot g^\inv$, $h(b_j)\mapsto g\cdot h(b_j) \cdot g^\inv$. As Poincar\'e transformations are isometries of Minkowski space, the spacetimes obtained by gluing these two regions are isometric and hence describe the same physical state. This motivates the appearance of the quotient by the action of $\pogr$ in formulas \eqref{eq:phase-space}, \eqref{eq:phase-space-with-holonomies}. Poincar\'e transformations which act diagonally on all holonomies have the status of gauge transformations: they are transitions between two different parametrisations of the same physical state.

\section{Gauge fixing and observers}

\subsection{The phase space of (2+1)-gravity as a constrained system}

Formula \eqref{eq:phase-space-with-holonomies} implies that the description of the phase space of (2+1)-gravity in terms of the auxiliary Poisson structure \eqref{eq:fr-bivector} can be viewed as a constrained system in the sense of Dirac. From this viewpoint, one has an extended phase space $\extphasespace=\pogr^{n+2g}$,  equipped with the Poisson structure \eqref{eq:fr-bivector}. The physical (or gauge-invariant) phase space is obtained from this extended phase space by imposing a set of $2n$ constraints which restrict the holonomies of the particles to the conjugacy classes $\grporbit$ and an $\pogr$-valued constraint which arises from the defining relation of the fundamental group $\pi_1(\surf)$. The former can be expressed as
\begin{equation}\label{ccond}
  \Tr(u_{M_i})-\cos\mu_i \weaklyequal 0, \qquad
  \Tr(j_{M_i}^aJ_a\cdot u_{M_i})- s_i\sin\mu_i\weaklyequal 0.
\end{equation}
The latter can be reformulated in terms of six constraints
\begin{equation}\label{eq:constraints-in-jp}
  \Tr(J_a\cdot u_C) \weaklyequal 0,\qquad  j_C^a\weaklyequal 0\qquad\forall a\in\{0,1,2\},
\end{equation}
where $(u_C,-\Ad(u_C)\bj_C)$ is the holonomy along the curve $c$ in Figure~\ref{fig:fundamental-group}:
\begin{equation}\label{groupconst}
  (u_C^\inv, \bj_C)
    \defeq M_1^\inv\cdots M_n^\inv [A_1^\inv, B_1]\cdots[A_g^\inv, B_g].
\end{equation}

It is shown in \cite{Meusburger:2003aa, Meusburger:2011aa} that the two constraints \eqref{ccond} associated with each particle are Casimir functions of the Poisson structure \eqref{eq:fr-bivector} and hence do not generate any gauge transformations.  In contrast, the six constraints \eqref{eq:constraints-in-jp} form a set of six first-class constraints. The associated gauge transformations these constraints generate via the Poisson bracket are the diagonal Poincar\'e transformations above which act on all holonomies by conjugation.

This implies that the description of the phase space in \eqref{eq:phase-space-with-holonomies} with its canonical Poisson structure can be obtained from Fock and Rosly's Poisson structure \eqref{eq:fr-bivector} on the ambient space $\extphasespace=\pogr^{n+2g}$ by imposing the $2n$ Casimir constraints \eqref{ccond} which implement the restriction of the holonomies $M_i$ to the conjugacy classes $\grporbit$ and the six first-class constraints \eqref{eq:constraints-in-jp} associated with the defining relation of the fundamental group $\pi_1(\surf)$.

\subsection{Constraints and gauge fixing conditions}

The description of the moduli space of flat connections in terms of the Poisson structure \eqref{eq:fr-bivector} has a direct geometrical interpretation and plays an important role as the starting point of the combinatorial quantisation formalism for Chern-Simons gauge theory. From the viewpoint of constrained systems, combinatorial quantisation is a ``constraint implementation after quantisation'' approach rooted in Dirac's constraint quantisation prescription. The formalism proceeds by first quantising the Poisson structure \eqref{eq:fr-bivector} on the ambient space and then imposing the quantum counterpart of the constraints \eqref{eq:constraints-in-jp} in the resulting quantum theory.

This formalism is well-established and proven to work for Chern-Simons theory with compact, semisimple gauge groups. In this case, the resulting quantum theory is given in terms of the representation theory of $q$-deformed universal enveloping algebras at roots of unity, and the implementation of the constraints is achieved via representation-theoretical methods. The formalism has been generalised to Chern-Simons theories with certain non-compact gauge groups \cite{Buffenoir:2002aa,Meusburger:2010bb}. However, there is no general formalism to treat the non-compact cases, because the representation theory of the associated quantum groups is more involved.

For this reason, it is desirable to also pursue other quantisation approaches which impose the constraints directly into the classical theory via gauge fixing and then attempt to quantise the resulting gauge-fixed theory. In the following, we apply the first step of this procedure to the moduli space of flat $\pogr$-connections. We gauge-fix the Poisson structure \eqref{eq:fr-bivector} on the ambient space $\extphasespace$ by imposing a set of six gauge fixing conditions associated with the constraints \eqref{eq:constraints-in-jp} following Dirac's gauge fixing procedure.

Dirac's gauge fixing procedure is a formalism that allows one to modify the Poisson structure of a constrained system in such a way that constraints and gauge fixing conditions are Casimir functions of this modified Poisson bracket, called the Dirac bracket.  We give a brief summary of this formalism for the case where all constraints are first-class and irreducible.

A constrained system with first-class constraints is a Poisson manifold $(M, \{\,,\,\})$ with constraint functions $\{\phi_i\}_{i=1,\dots,k}\subset\cif(\extphasespace)$ such that the Poisson bracket of two constraint functions vanishes on the constraint surface $\bigcap_{i=1}^k \phi_i^\inv(0)\subset M$. Gauge fixing consists in imposing an additional set of constraints $\{\chi_j\}_{j=1,\dots,k}\subset\cif(\extphasespace)$, the gauge fixing conditions, such that the following two requirements are met:
\begin{enumerate}
\item It is possible to map any point $p\in \bigcap_{i=1}^k \phi_i^\inv(0)$ on the constraint surface to one that satisfies the gauge fixing conditions via the flows on $M$ generated by the constraint functions $\phi_i$.
\item The gauge fixing conditions must break the gauge symmetries completely. In other words, the matrix $C=(\{\phi_i,\chi_j\})_{i,j=1,\dots,k}$ must be invertible at least on the gauge-fixed constraint surface $\csurface \defeq \bigcap_{i=1}^k \phi_i^\inv(0) \cap \chi_i^\inv(0)$.
\end{enumerate}
If these conditions are satisfied, the Dirac matrix $D=(\{C_i,C_j\})_{i,j=1,\dots,2k}$ obtained by combining the constraints $\phi_i$ and the gauge fixing conditions $\chi_j$ into a set of $2k$ constraints $\{C_i\}_{i=1,\dots,2k}$ is invertible on $\csurface$. The Dirac bracket $\{\,,\}_D$ is given in terms of its inverse as
\begin{equation*}
  \{F, G\}_D \defeq \{F, G\} - \smashoperator{\sum_{i,j=1}^{2k}} \{F, C_i\} (D^\inv)_{ij} \{C_j, G\}\qquad\forall F,G\in\cif(M).
\end{equation*}
and defines a Poisson structure on $\csurface$ for which the constraints $\phi_i$ and gauge fixing conditions $\chi_i$ are Casimir functions. This allows one to strongly impose these constraints in the classical theory.

In the application to the phase space of (2+1)-gravity, the Poisson manifold $(M,\{\,,\})$ is the manifold $\extphasespace=\pogr^{n+2g}$ with the Poisson bracket \eqref{eq:fr-bivector} and the constraint functions $\phi_i$ are given by \eqref{eq:constraints-in-jp}. The associated gauge transformations these constraints generate via the Poisson bracket correspond to the diagonal action of $\pogr$ on $\pogr^{n+2g}$ and encode the gauge freedom of applying a Poincar\'e transformation to the regions in Minkowski space from which spacetimes are obtained by gluing.

This gauge freedom is linked to the absence of a preferred observer in general relativity. Physical measurements of quantities such as the distance between certain particles or quantities associated with the geometry of the handles on $\surf$ depend on the choice of the observer. If one restricts attention to observers in free fall, each observer corresponds to a timelike geodesic in Minkowski space and the Poincar\'e transformations relating different domains in Minkowski space describe  the transitions between different observers. Eliminating this gauge freedom via gauge fixing conditions thus corresponds to specifying an observer.

In the absence of a preferred reference frame, the only physically meaningful way of specifying an observer is with respect to the geometry of the spacetime itself, for instance with respect to the geometry of a handle or two point particles contained in the spacetime. A detailed discussion of this issue  is given in \cite{Meusburger:2011aa} and in \cite{Meusburger:2009aa} for the case of vacuum spacetimes.

In the following, we will restrict attention to gauge fixing conditions based on the motion of two point particles. As permutations of the particles correspond to the action of the braid groups on $\surf$ and the associated surface $\surf\setminus D$ with a disc removed and these braid groups act by Poisson isomorphisms \cite{Meusburger:2003aa}, we can suppose without restriction of generality that these point particles are the ones associated with the holonomies $M_1,M_2$.

\begin{figure}
  \centering
  \includegraphics{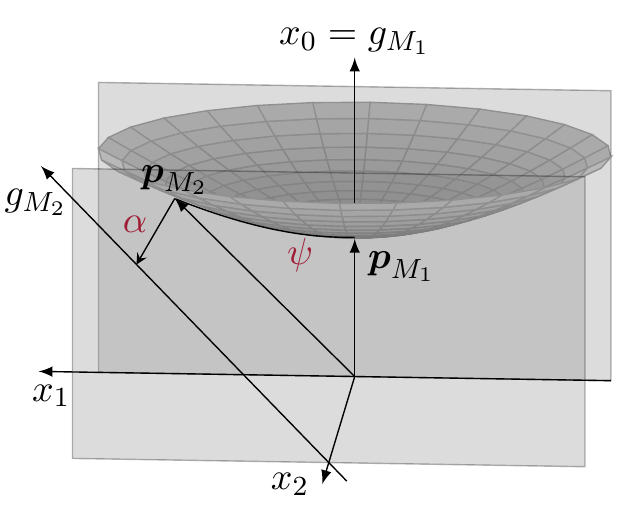}
  \caption{The two dynamical variables $\psi, \alpha$ characterising the relative motion of two point particles.}
  \label{fig:2partgfix}
\end{figure}

A set of particularly simple gauge fixing condition of this type is investigated in \cite{Meusburger:2011aa}. These gauge fixing conditions impose that the first particle is at rest at the origin and that the second particle moves in the direction of the $x_1$-axis in such a way that its distance to the first particle is minimal at the intersection point of the two worldlines with the $x_1x_2$-plane. These conditions on the particles' worldlines are depicted in Figure~\ref{fig:2partgfix}.

After such a gauge fixing condition is imposed, only two of the original eight degrees of freedom associated with the two particles remain: their relative velocity $v=\tanh \psi$ and their minimal distance $\alpha$, as indicated in Figure~\ref{fig:2partgfix}. These two quantities give a Poincar\'e-invariant characterisation of the resulting two-particle system and are given as conjugation-invariant functions of the product $M_2\cdot M_1$ of their holonomies. The variable $\psi$ depends only on its Lorentzian component while $\alpha$ involves both, its Lorentzian and translational component, and is linear in the latter.

Although these gauge fixing conditions from \cite{Meusburger:2011aa} and the parametrisation of the two Poincar\'e-invariant degrees of freedom of the two-particle system are motivated by their direct physical interpretation, they are not unique. In the following, we therefore consider more general gauge fixing conditions which are subject to the following two structural requirements:
\begin{enumerate}
\item The gauge fixing conditions are functions of the holonomies $M_1,M_2$.
\item The gauge fixing conditions involve three conditions that depend only on the Lorentzian components of $M_1,M_2$ and three conditions that are linear in the variables $\bj_{M_1}$, $\bj_{M_2}$:
\begin{equation}\label{eq:gfix-conditions}
  \sum_{i=1}^2 \Theta^{M_i}_{a}\,j^a_{M_i} \weaklyequal 0, \quad
  \sum_{i=1}^2 \Gamma^{M_i}_{a}\,j^a_{M_i} \weaklyequal 0, \quad
  \sum_{i=1}^2 \Omega^{M_i}_{a}\,j^a_{M_i} \weaklyequal 0, \quad
  \Delta_1 \weaklyequal 0, \quad
  \Delta_2 \weaklyequal 0, \quad
  \Delta_3 \weaklyequal 0,
\end{equation}
where $\Theta^{M_i}_{a},\Gamma^{M_i}_{a}, \Omega^{M_i}_{a}, \Delta_j \in \cif(\logr\times\logr)$ and the two copies of the Lorentz group $\logr$ are identified with the Lorentzian components of the holonomies $M_1$ and $M_2$
\end{enumerate}
Both conditions are well-motivated. The first condition ensures that the gauge fixing conditions have a direct physical interpretation by specifying an observer with respect to the motion of two particles. The second is motivated by the wish to preserve the canonical $\mathbb N$-grading of the Poisson structure \eqref{eq:fr-bivector} which is associated with a physical dimension and plays an important role in the quantisation of the theory.

We also allow more freedom in the parametrisation of the Poincar\'e-invariant degrees of freedom of the two-particle system by admitting general Poincar\'e-invariant functions of the product $M_2\cdot M_1=(u_{12}, -\Ad(u_{12})\bj_{12})$ defined as follows:
\begin{align}\label{eq:psi-alpha-general}
  \psi=f(\Tr(u_{12})), \quad
  \alpha=g(\Tr(u_{12})) \Tr(j_{12}^aJ_a\cdot u_{12})+h(\Tr(u_{12})),
\end{align}
with diffeomorphisms $f,g\in\cif(\RR)$ and a smooth function $h\in\cif(\RR)$. As before, the variable $\psi$ characterises the Lorentzian component of the group element $M_2\cdot M_1$ and $\alpha$ depends on both, its Lorentzian and translational component. It follows directly from the cyclic invariance of the trace that both quantities are Poincar\'e-invariant and hence independent of the choice of observer.

\section{The Dirac bracket}

As the phase space of (2+1)-gravity is given as a constrained system with six first-class constraints, the construction of the associated Dirac bracket involves inverting a ($12\times 12$)-Dirac matrix. This could lead one to expect that the calculation of this Dirac bracket is not feasible or that the resulting Dirac bracket takes a very complicated form. However, it turns out that this is not the case. The Dirac bracket associated with the constraints \eqref{eq:constraints-in-jp} and general gauge fixing conditions of the form \eqref{eq:gfix-conditions} is derived in \cite{Meusburger:2012aa}.  It is shown there that it extends canonically to a Poisson structure on $\RR^2\times \pogr^{n+2g-2}$, where $\RR^2$ is parametrised by the variables $\psi,\alpha$ from \eqref{eq:psi-alpha-general} and $\pogr^{n+2g-2}$ by the non-gauge-fixed holonomies $M_3,\dots, M_n, A_1,B_1,\dots,A_g,B_g$.

\begin{theorem}[\cite{Meusburger:2012aa}]\label{thm:generic-dirac-bracket}
  The constraints \eqref{eq:constraints-in-jp} and gauge fixing conditions \eqref{eq:gfix-conditions} define a Poisson structure $\{\,,\,\}_D$ on the gauge-fixed constraint surface $\Sigma$ that extends to a bracket on $\RR^2\times \pogr^{n+2g-2}$. This bracket takes the following form:
  \begin{enumerate}
  \item The Dirac bracket of $\psi$ and $\alpha$ vanishes: $\{\psi, \alpha\}_D = 0$, and for $X \in \{M_3, \dots, B_g\}$ and $f\in\cif(\logr^{n+2g-2})$:
    \begin{equation*}
      \begin{aligned}
        \{\psi, f \}_D &= 0, &
        \{\psi, \bj_X \}_D &= -\idadi{X} \, \bq_\psi, \\
        \{\alpha, f\}_D &= \smashoperator{\sum_{Y\in\{M_3,\dots,B_g\}}} q_\alpha^a(J_a^{R,Y}+J_a^{L,Y})f, &
        \{\alpha, \bj_X\}_D &= -\idadi{X} \bq_\theta - \bq_\alpha \wedge \bj_X,
      \end{aligned}
    \end{equation*}
    with $\bq_\psi, \bq_\alpha,\bq_\theta: \RR^2 \to \RR^3$ satisfying $\bq_\psi \wedge \bq_\alpha = 0$ and $\partial_\alpha\bq_\psi = \partial_\alpha\bq_\alpha = \partial_\alpha^2\bq_\theta= 0$.
  \item For $F, G \in \cif(\nongaugephasespace)$ we have $\{F, G\}_D = B_{\mathrm{FR}}^r(\diffd F \otimes \diffd G)$, where $B_{\text{FR}}^r$ is the Poisson bivector \eqref{eq:fr-bivector} without the holonomies $M_1,M_2$ and $r: \RR^2 \to \poal \oo \poal$ is of the form
    \begin{equation*}
      r (\psi,\alpha) = P_a \otimes J^a - V^{bc}(\psi)(P_b \otimes J_c - J_c \otimes P_b) + \ee^{bcd}m_d(\psi,\alpha) P_b \otimes P_c,
    \end{equation*}
    where $V:\RR \to \Mat(3, \RR)$ and $\bm:\RR^2 \to \RR^3$ satisfies $\partial_\alpha^2 \bm=0$.
  \end{enumerate}
\end{theorem}

As is apparent from this theorem, the resulting Dirac bracket has a very simple form and is closely related to the original Poisson structure \eqref{eq:fr-bivector} on the extended phase space $\extphasespace$. The Poisson brackets of functions of the non-gauge-fixed holonomies are again given by the Poisson bivector \eqref{eq:fr-bivector}. The only difference is that the holonomies of the two gauge-fixed particles are removed from the description and instead of a classical $r$-matrix, this Poisson bivector is now determined by a map $r:\RR^2\to\poal\oo\poal$ whose arguments are the two Poincar\'e-invariant variables $\psi,\alpha$ associated with the gauge-fixed particles. The variables $\psi$ and $\alpha$ Poisson-commute, and their Poisson brackets with functions of the residual holonomies are given by $\bq_\psi,\bq_\alpha,\bq_\theta: \RR^2\to\RR^3$.

It is shown in \cite{Meusburger:2012aa} that the vector $\bq_\psi$ is closely related to the total momentum $\bp_{12}$ of the two gauge-fixed particles and $\bq_\theta$ to their angular momentum $\bj_{12}$ given as in \eqref{mpartpar}, \eqref{ujexpress} as a function of  $M_2\cdot M_1$. The group element $M_2\cdot M_1$ defines a cone in Minkowski space whose axis is the unique geodesic that is stabilised by $M_2\cdot M_1$ and whose deficit angle and time shift are determined by, respectively, $\psi$ and $\alpha$. However, unlike in the case of point particles, the Lorentzian component $u_{12}\in \logr$ of the holonomy $M_2\cdot M_1$ is not required to be elliptic, but can become parabolic or hyperbolic. This implies that the axis of the associated cone can also become a light- or spacelike geodesic. This occurs when the relative velocity of the two gauge-fixed particles becomes large and corresponds to the formation of Gott pairs \cite{Gott:1991aa}.

The other main difference is that the mass and spin variables associated with $M_2\cdot M_1$ via \eqref{ccond} are no longer fixed parameters but given as functions of the variables $\psi$ and $\alpha$. The deficit angle and time shift associated to this cone are therefore dynamical: they depend on the relative velocity and minimal distance of the two gauge-fixed particles. The former is determined by the parameter $\psi$ and encoded in the Lorentzian component of $M_2\cdot M_1$, the latter by $\alpha$ and involves also the translational component.

\begin{figure}
  \centering
  \includegraphics{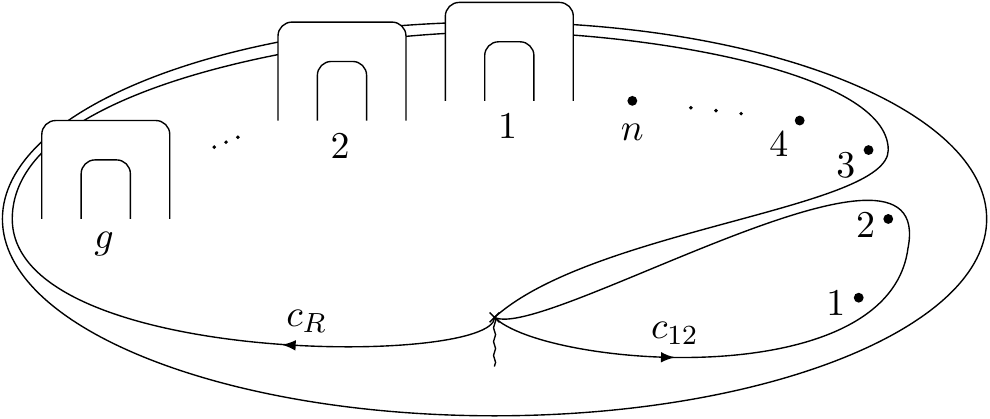}
  \caption{The homotopic curves $c_{12}$ and $c_R$ corresponding to the holonomies $M_{12} = M_2 \cdot M_1$ and $M_R = M_3^\inv \cdots M_n^\inv \cdot [A_1^\inv,B_1] \cdots [A_g^\inv, B_g]$.}
  \label{fig:fundamental-group-12-R}
\end{figure}

By imposing the constraints \eqref{groupconst}, one identifies the Poincar\'e element $M_2\cdot M_1$ that characterises this cone with the holonomy of a curve around all non-gauge-fixed particles and handles, as shown in Figure~\ref{fig:fundamental-group-12-R}:
\begin{equation*}
  M_2 \cdot M_1 \weaklyequal M_3^\inv \cdots M_n^\inv \cdot [A_1^\inv,B_1] \cdots [A_g^\inv, B_g]\in\pogr.
\end{equation*}
This amounts to a description of the centre of mass of the residual system defined by the non-gauge-fixed particles and handles as an effective particle. The geodesic stabilised by the Poincar\'e element $M_2\cdot M_1$ describes the motion of the centre of mass with respect to the observer determined by the two gauge-fixed particles. The two parameters $\psi$, $\alpha$ determine, respectively, its energy and angular momentum as measured by this observer.

\subsection{Classical dynamical r-matrices}

The simple form of the Dirac bracket and its close relation to the Poisson bracket \eqref{eq:fr-bivector} on the extended phase space suggest that there is an underlying mathematical structure which ensures the consistency of this description and implies that the Dirac bracket  satisfies the Jacobi identity. For the original bracket \eqref{eq:fr-bivector} on the extended phase space $\extphasespace$, this is ensured by the classical Yang-Baxter equation. It is shown in \cite{Meusburger:2012aa} that a generalisation of this result also holds for the Dirac bracket. The main difference is that the relevant mathematical structure is no longer a  classical $r$-matrix, \ie a solution of the classical Yang-Baxter equation,  but a classical dynamical $r$-matrix, which solves the classical dynamical Yang-Baxter equation.

\begin{theorem}[\cite{Meusburger:2012aa}]
The bracket from Theorem~\ref{thm:generic-dirac-bracket} satisfies the Jacobi identity if and only if the function $r:\RR^2\to \poal\oo \poal$ is a solution of the classical dynamical Yang-Baxter equation 
\begin{multline*}
[r_{12}, r_{13}]+[r_{12}, r_{23}]+[r_{13}, r_{23}]=\\
        x_\alpha^{(1)}\partial_{\alpha}\,r_{23} - x_\alpha^{(2)}\partial_{\alpha}\,r_{13} + x_\alpha^{(3)}\partial_{\alpha}\,r_{12}
      + x_\psi^{(1)}\partial_{\psi}\,r_{23} - x_\psi^{(2)}\partial_{\psi}\,r_{13} + x_\psi^{(3)}\partial_{\psi}\,r_{12},
\end{multline*}
with $x_\alpha=q_\alpha^a J_a+q_\theta^a P_a$, $x_\psi=q_\psi^a P_a$, and  of the following additional equations:
\begin{equation}\label{eq:q-relations}
  \left.
  \begin{aligned}
    0 &= q_\psi^a + \tensor{\ee}{^a_b_c} q_\psi^b \partial_\psi q_\psi^c + q_\psi^b \tensor{V}{_b^a} - q_\psi^a \tensor{V}{^b_b}, \\
    0 &= \tensor{\ee}{^a_d_h} q_\alpha^d V^{bh} + \tensor{\ee}{^b_d_h} q_\alpha^d V^{ah} + \tensor{\ee}{_c_d_e} q_\alpha^c V^{de} \eta^{ab} - \tensor{\ee}{^b_d_e} q_\alpha^a V^{de} + q_\alpha^a \partial_\alpha q_\theta^b - q_\psi^b \partial_\psi q_\alpha^a, \\
    0 &= q_\theta^a + \tensor{\ee}{^a_b_c} q_\theta^b \partial_\alpha q_\theta^c + \tensor{\ee}{^a_b_c} q_\psi^b \partial_\psi q_\theta^c - \tensor{\ee}{^a_b_c} m^b q_\alpha^c + q_\theta^d \tensor{V}{_d^a} - q_\theta^a \tensor{V}{_d^d}.
  \end{aligned}
  \quad\right\}
\end{equation}
\end{theorem}
It is shown in \cite{Meusburger:2012aa} that the classical dynamical Yang-Baxter equation (CDYBE) guarantees the Jacobi identity for Poisson brackets involving functions of the residual non-gauge-fixed holonomies.  The additional conditions \eqref{eq:q-relations} ensure that the Jacobi identity also holds for mixed brackets which involve both, functions of the non-gauge-fixed holonomies, and functions of the variables $\psi,\alpha$.  The classical dynamical Yang-Baxter and the additional conditions \eqref{eq:q-relations} thus take the place of the classical Yang-Baxter equation in the Poisson structure \eqref{eq:fr-bivector}, and the appearance of classical dynamical $r$-matrices is related to the implementation of an observer into the description.  The two variables arising in the classical dynamical $r$-matrix have a direct physical interpretation: they correspond to the total energy and angular momentum of the spacetime as measured by this observer.

A similar pattern was found in \cite{Buffenoir:2003aa, Buffenoir:2005aa, Noui:2005aa}, where a regularisation procedure for point particles coupled to $\SL(2,\CC)$-Chern-Simons theory lead to a description involving classical dynamical $r$-matrices and observers. As the formalism and description used in these works are very different and do not involve gauge fixing the Poisson structure \eqref{eq:fr-bivector}, these results suggests that the appearance of classical dynamical $r$-matrix symmetries together with observers is not limited to specific models or gauge fixing procedures but a generic feature of (2+1)-gravity.

This has important implications for the quantisation of the theory and for the question which quantum groups are relevant to quantum gravity in (2+1)-dimensions, which has been the subject of much debate. As the Poisson-Lie symmetries associated with the Poisson structure \eqref{eq:fr-bivector} can be viewed as a classical counterpart or first-order approximation of the quantum group symmetries in the associated quantum theory, the classical (dynamical) $r$-matrices arising in this description allow one to draw conclusions about the relevant quantum group.

While some results \cite{Meusburger:2010bb,Meusburger:2010aa,Meusburger:2009ab} suggest that the Drinfel'd double $\mathcal D(\logr)$ of the three-dimen\-sional Lorentz group is the relevant quantum group for the quantisation of the Poisson structure \eqref{eq:fr-bivector} on the extended phase space, the results of the gauge fixing suggest that the implementation of an observer in the resulting quantum theory leads to the appearance of {\em dynamical} quantum groups. The dynamical variables of these dynamical quantum groups should be related to the total energy and angular momentum of the universe as measured by this observer. This has interesting implications for the physical interpretation of the theory and suggests that the  role of quantum group symmetries in (2+1)- and higher-dimensional gravity is more subtle than apparent at first sight.

\subsection{Dynamical Poincar\'e transformations and the centre-of-mass frame}

The choice of gauge fixing conditions in \cite{Meusburger:2011aa} is particularly simple and motivated by its direct physical interpretation. However, it is obvious that this condition is not unique or distinguished from similar conditions imposed on the motion of the two particles. This implies that there should be dynamical transformations which depend on the variables $\psi,\alpha$ and relate different gauge choices.

Each gauge choice leads to a cone determined uniquely by the product $M_2\cdot M_1$ of the holonomies of the two gauge-fixed particles and such cones can be related by Poincar\'e transformations. This suggests that the dynamical transformations should be Poincar\'e transformations which depend on the variables $\psi, \alpha$. Our requirements for the gauge fixing conditions imply that the Lorentzian components of these dynamical Poincar\'e transformations should depend only on $\psi$ while their translational component should depend on both $\psi$ and $\alpha$, but on $\alpha$ at most linearly.

In this section we will show that dynamical Poincar\'e transformations $p=(g,-\Ad(g)\bt)\in\cif(\RR^2,\pogr)$ with $\partial_\alpha g=\partial_\alpha^2\bt=0$ can be interpreted as transformations of the classical dynamical $r$-matrices and allow one to locally relate each Poisson structure obtained from gauge fixing to a particularly simple standard solution. As a first step, we determine the transformation of the Dirac brackets under dynamical Poincar\'e transformations.

\begin{lemma}[\cite{Meusburger:2012aa}]\label{lem:poinc-trafo}
Let $\{,\}_D$ and $r: \RR^2 \to \poal \oo \poal$ be given as in Theorem~\ref{thm:generic-dirac-bracket} and consider a Poincar\'e transformation $p$ as above that acts on the residual holonomies $M_3,\dots,B_g$ by conjugation:
\begin{equation*}
  \Phi^p:\
  (\psi,\alpha,M_3,\dots,B_g)\mapsto
  \bigl(\psi,\alpha,\; p(\psi,\alpha)\cdot M_3\cdot p(\psi,\alpha)^\inv,\dots,\;p(\psi,\alpha)\cdot B_g\cdot p(\psi,\alpha)^\inv\bigr).
\end{equation*}
Then for all $F,G\in\cif(\RR^2\times \pogr^{n+2g-2})$: $$\{ F\circ \Phi^p, G\circ \Phi^p\}_D=\{F,G\}^p_D\circ \Phi^p,$$ where $\{\,,\,\}_D^p$ is the bracket from Theorem~\ref{thm:generic-dirac-bracket} associated with
\begin{equation}\label{eq:trafo-quant}
  \left.
  \begin{aligned}
    &\bq_\psi^p=\Ad(g)\bq_\psi,\qquad \bq_\alpha^p=\Ad(g)\bq_\alpha,\qquad \bq_\theta^p=\Ad(g)(\bq_\theta-\bq_\alpha\wedge\bt),\\
    &r^p=\left(\Ad(p)\oo\Ad(p)\right)\left[r+(q^a_\alpha J_a+q^a_\theta P_a)\wedge p^\inv \partial_\alpha p+q^a_\psi P_a\wedge p^\inv \partial_\psi p\right].
  \end{aligned}
  \qquad\right\}
\end{equation}
\end{lemma}

This lemma shows that the dynamical Poincar\'e transformations relating different gauge choices can be identified with a simultaneous transformation of the classical dynamical $r$-matrix $r:\RR^2\to\poal\oo\poal$ and the vector-valued maps $\bq_\psi,\bq_\alpha,\bq_\theta:\RR^2\to\RR^3$ in Theorem~\ref{thm:generic-dirac-bracket}. In particular, this implies that the Poincar\'e-transformed quantities $r^p,\bq_\psi^p,\bq_\alpha^p,\bq_\theta^p$ satisfy the CDYBE and the additional conditions \eqref{eq:q-relations} if and only if the original quantities $r,\bq_\psi,\bq_\alpha,\bq_\theta$ satisfy them.

\begin{corollary}\label{cor:rmat}
Let $r: \RR^2\to \poal \oo \poal$, $\bq_\psi,\bq_\alpha,\bq_\theta:\RR^2\to\RR^3$ as in Theorem~\ref{thm:generic-dirac-bracket} be a solution of the CDYBE that satisfies the conditions in \eqref{eq:q-relations} and let $p:\RR^2\to \pogr$ be a dynamical Poincar\'e transformation as in Lemma~\ref{lem:poinc-trafo}. Then $r^p:\RR^2\to\poal\oo\poal$, $\bq_\psi^p,\bq_\alpha^p,\bq_\theta^p:\RR^2\to\RR^3$ given by \eqref{eq:trafo-quant} are solutions of the CDYBE and conditions \eqref{eq:q-relations}, and the map $\Phi^p$ defines a Poisson isomorphism between the Poisson structures $\{\,,\,\}_D$ and $\{\,,\,\}_D^p$.
\end{corollary}

This corollary implies that these dynamical Poincar\'e transformations can be viewed as a generalisation of the usual gauge transformations of classical dynamical $r$-matrices introduced in \cite{Etingof:1998aa}. The difference is that in our case, these classical dynamical $r$-matrices are not required to be invariant under the action of a fixed Cartan subalgebra, but are associated with a two-dimensional subalgebra of $\poal$ which is defined by $\bq_\psi,\bq_\alpha,\bq_\theta$ and allowed to vary with the variables $\psi,\alpha$. It is shown in \cite{Meusburger:2012aa} that the conditions \eqref{eq:q-relations} can be viewed as a generalisation of the invariance condition in \cite{Etingof:1998aa} to this setting.

In view of Corollary~\ref{cor:rmat}, it is natural to ask if by applying such dynamical Poincar\'e transformations to solutions of the CDYBE, it is possible to relate them to a set of particularly simple standard solutions. As the classical dynamical $r$-matrices together with the maps $\bq_\psi,\bq_\alpha,\bq_\theta:\RR^2\to\RR^3$ determine the Dirac bracket completely, this would amount to a complete classification of the Poisson structures resulting from our gauge fixing conditions.

It is shown in \cite{Meusburger:2012aa} that this is indeed possible for those values of the variable $\psi$ for which $\bq_\psi(\psi)$ and $\bq_\alpha(\psi)$ are time- or spacelike if, additionally, one performs a suitable rescaling of the parameters $\psi$ and $\alpha$:

\begin{theorem}[\cite{Meusburger:2012aa}]\label{thm:dcybe-standard-transformation}
Let $I\subset \RR$ be an open interval and $r:I\times \RR\to\poal\oo\poal$, $\bq_\psi,\bq_\alpha,\bq_\theta:\RR^2\to \RR^3$  a solution of the CDYBE  and of equations \eqref{eq:q-relations} for which $\bq_\psi^2,\bq_\alpha^2\neq 0$, $\bq_\psi\wedge\bq_\alpha=0$ and $\partial_\alpha \bq_\psi = \partial_\alpha \bq_\alpha = \partial_\alpha^2 \bq_\theta = 0$ on $I\times\RR$. Then there exists a Poincar\'e transformation $p:I\times\RR\to \pogr$  and a diffeomorphism  $\by=(y^1, y^2): I\times\RR\to I'\times\RR$ with $\partial_\alpha y_1=\partial_\alpha^2 y_2=0$ such that one of the following holds:
\begin{enumerate}
\item $\bq_\psi^p,\bq_\alpha^p,\bq_\theta^p \in \Span\{\be_0\}$ for all $\psi \in I$ and
\begin{equation*}
  r^p(\psi,\alpha)=\tfrac 1 2 (P_a\!\oo\! J^a\!+\!J^a\!\oo\! P_a)\!+\!\tfrac 1 2 \tan\tfrac {y^1(\psi)} 2 \left(P_1\!\wedge\! J_2\!-\!P_2\!\wedge\! J_1\right)\!+\!\frac{y^2(\psi,\alpha)}{4\cos^2\tfrac {y^1(\psi)} 2}P_1\!\wedge\! P_2,
\end{equation*}

\item $\bq_\psi^p,\bq_\alpha^p,\bq_\theta^p \in \Span\{\be_1\}$ for all $\psi\in I$ and
\begin{equation*}
  r^p(\psi,\alpha)=\tfrac 1 2 (P_a\!\oo\! J^a\!+\!J^a\!\oo\! P_a)\!+\!\tfrac 1 2 \tanh\tfrac {y^1(\psi)} 2 \left(P_2\!\wedge\! J_0\!-\!P_0\wedge J_2\right)\!+\!\frac{y^2(\psi,\alpha)}{4\cosh^2\tfrac {y^1(\psi)} 2}P_2\!\wedge\! P_0.
\end{equation*}
These classical dynamical $r$-matrices are invariant under, respectively, the action of the Cartan subalgebras $\ah_1=\Span\{J_0,P_0\}$ and $\ah_2=\Span\{J_1,P_1\}$.
\end{enumerate}
\end{theorem}

This result is intuitive from the perspective of Lie algebras, as every Cartan subalgebra of $\poal$ is conjugate to either $\ah_1$ or $\ah_2$. Moreover, it has a direct geometrical interpretation: As discussed in the previous sections, the product $M_2\cdot M_1$ of the two gauge-fixed holonomies determines a cone in Minkowski space whose axis is the geodesic stabilised by $M_2\cdot M_1$. The direction of its axis is given by $\bq_\psi$ and its offset orthogonal to its axis is encoded in $\bq_\theta$. If $\bq_\psi$ is timelike or spacelike, the associated geodesic can be mapped to, respectively, the $x_0$-axis or the $x_1$-axis via a suitable Poincar\'e transformation that depends on the variables $\psi,\alpha$. This transforms the associated classical dynamical $r$-matrix and brings $\bq_\psi,\bq_\alpha,\bq_\theta$ into the form in Theorem~\ref{thm:dcybe-standard-transformation}. The role of the rescaling in Theorem~\ref{thm:dcybe-standard-transformation} is to eliminate the freedom in defining $\psi,\alpha$ as functions of the Lorentzian and translational component of the holonomy $M_2\cdot M_1$ in \eqref{eq:psi-alpha-general} and defines the mass and spin of the associated cone as a function of $y_1$ and $y_2$.

From the discussion in the previous sections it then follows that the timelike solution in Theorem~\ref{thm:dcybe-standard-transformation} corresponds to an observer in the centre-of-mass frame of the universe, to whom the centre of mass appears as a particle that is at rest at the origin. The interpretation of the spacelike solution in Theorem~\ref{thm:dcybe-standard-transformation} is less direct. It arises for those values of the parameter $\psi$ for which the gauge-fixed holonomies $M_1,M_2$ form a Gott pair \cite{Gott:1991aa} and can be viewed as a ``tachyonic'' particle whose worldline is identified with the $x_1$-axis.

This amounts to a complete {\em local} classification of all possible Dirac brackets for those values of the parameter $\psi$ for which $\bq_\psi^2(\psi), \bq_\alpha^2(\psi)\neq 0$. Theorem~\ref{thm:dcybe-standard-transformation} states that every such solution can be identified with one of the two standard solutions above for all values of the parameter $\psi$ for which $\bq_\psi^2(\psi), \bq_\alpha^2(\psi)\neq 0$. The possible outcomes of the gauge fixing procedure are thus equivalent to the ones associated with the centre-of-mass frames, where the centre of mass of the residual system appears as a particle at rest at the origin or a ``tachyonic'' particle associated with the $x_1$-axis.

However, it is shown in \cite{Meusburger:2011aa, Meusburger:2012aa} that for generic Dirac brackets resulting from our gauge fixing conditions the signature of $\bq_\psi, \bq_\alpha$ varies with $\psi$. In particular,  the solutions described there exhibit values of the variable $\psi$ for which these vectors become lightlike. These lightlike solutions appear as transition points between the time- and spacelike cases discussed above and do not correspond to a fixed Cartan subalgebra of $\poal$. The solutions of the CDYBE resulting from gauge fixing in (2+1)-gravity are therefore not simply given by standard classical dynamical $r$-matrices as in \cite{Etingof:1998aa} but connect in a non-trivial way non-equivalent standard classical dynamical $r$-matrices associated with non-conjugate Cartan subalgebras of $\iso(2,1)$.

\section{Outlook and conclusions}

In this article we investigated gauge fixing in (2+1)-dimensional gravity with vanishing cosmological constant. We discussed how gauge fixing this theory amounts to the introduction of an observer and determined the resulting Dirac bracket for a rather general set of gauge fixing conditions. We showed that these Dirac brackets are determined by solutions of the classical dynamical Yang-Baxter equation, whose two dynamical variables have a direct interpretation as the total energy and total angular momentum of the spacetime measured by this observer.  We showed how dynamical Poincar\'e transformations allow one to relate, for almost all values of the dynamical variables, any Dirac bracket obtained from this gauge fixing to two particularly simple standard solutions which correspond to the centre-of-mass frame of the spacetime.

That these statements hold for a very large class of gauge fixing conditions and that classical dynamical $r$-matrix symmetries were also obtained in a different and independent approach based on the regularisation of punctures coupled to Chern-Simons theory \cite{Buffenoir:2003aa, Buffenoir:2005aa, Noui:2005aa} suggest that the appearance of classical dynamical $r$-matrices in gauge-fixed (2+1)-gravity is generic and related to the introduction of observers into the theory.

From the physics perspective, this result suggests that the relevant quantum groups for the quantisation of (2+1)-gravity should become dynamical quantum groups when an observer is implemented in the quantum theory.  As the dynamical variables in the classical theory are the total energy and angular momentum measured by this observer, the dynamical quantum group symmetries, unlike the quantum groups acting on extended, non-gauge-invariant Hilbert spaces, should manifest themselves in the measurements of  observers.

It is also instructive to compare the classical dynamical $r$-matrices arising in this description to the Hopf algebra symmetries associated with other models such as $\kappa$-Poincar\'e symmetries.  By comparing the $r$-matrices appearing in these descriptions, one finds that the classical dynamical $r$-matrices in Theorem~\ref{thm:dcybe-standard-transformation} are rather similar to the ones occurring in (2+1)-dimensional $\kappa$-Poincar\'e symmetries but do not coincide with them. The difference is that the classical dynamical $r$-matrices  in this article also contain a term which involves the tensor product of two generators of translations, which is not present in the classical $r$-matrices of $\kappa$-Poincar\'e symmetries. It would be interesting to investigate if $\kappa$-Poincar\'e symmetries can be obtained from the classical dynamical $r$-matrices in Theorem~\ref{thm:dcybe-standard-transformation} via a suitable limiting procedure in which the angular momentum of the cone tends to zero.

Finally, it would be desirable to quantise the gauge-fixed Poisson structure derived in this article. As the Dirac brackets are of a form very similar to the Poisson structure in \cite{Fock:1998aa, Alekseev:1995ab} which serves as the starting point of the combinatorial quantisation formalism, this formalism could also be applied to quantise the gauge-fixed theory. In that case the difficulties associated with the implementation of the constraints would no longer be present and the quantisation of the theory would reduce to the construction of the dynamical quantum groups that correspond to the classical dynamical $r$-matrices in the classical description. In particular, it would be interesting to see if this construction gives rise to quantisation conditions on the total mass and angular momentum of the spacetime which are encoded in its dynamical variables.

\bibliographystyle{custom}
\bibliography{corfu_2011}{}

\end{document}